\documentclass{appolb-yost}
\usepackage{epsfig}
\usepackage{cite}
\newcommand {\Sp}{{\rm Sp }}
\headauthor{S.A. Yost, S. Jadach, B.F.L. Ward}
\headtitle{Radiative Corrections to Bremsstrahlung in Radiative Return}
\newcommand{\pptnum}{BU-HEPP-05-02}
\Preprintfalse
\Redakcjatrue

\begin{document}
\title{Radiative Corrections to Bremsstrahlung in Radiative Return%
\thanks{Presented at the Cracow Epiphany Conference on Hadron Spectroscopy,
5 -- 8 Jan., 2005}%
}
\author{Scott A. Yost
\address{Department of Physics, Baylor University, Waco, Texas 76798, USA}
\\[12pt]
S. Jadach
\address{Institute of Nuclear Physics, ul. Radzikowskiego 152, Krak\'ow, Poland\\
CERN, Theory Division, CH-1211, Geneva 23, Switzerland}
\and
B.F.L. Ward
\address{Department of Physics, Baylor University, Waco, Texas 76798, USA}
}
\maketitle
\begin{abstract}
Radiating a photon from the initial state provides a useful tool for 
studying a range of low energy physics using a high-energy $e^+ e^-$ 
accelerator. Accurate results require careful calculation of the first 
order virtual photon corrections. We compare exact results for initial 
state radiative corrections, finding agreement to an order of $10^{-5}$ or 
better as a fraction of the Born cross-section for most of the range of 
photon energies, at CMS energies relevant in both high-energy collision and 
radiative return experiments.
\end{abstract}
\PACS{12.15.Lk, 13.66.De}
  
\section{Introduction}

Radiative return\cite{rad1,rad2,rad3} provides a mechanism 
for exploring a wide range of 
CMS energies in hadron production in a high luminosity $e^+e^-$ collider.
Radiating a hard photon from the initial state (ISR) reduces the effective
energy of the collision, allowing a range of energies to be scanned by 
observing different values of the hard photon energy. A precise calculation
of this process requires including the ${\cal O}(\alpha^2)$ contributions
arising from an additional virtual or soft photon.  These effects were 
integrated into a MC generator PHOKHARA designed to calculate radiative 
return at DA$\Phi$NE, CLEO-C and $B$ factories.\cite{rad4,rad5,rad6} 
The PHOKHARA MC is discussed by H. K\"uhn in these proceedings.\cite{kuhn}
Both the processes $e^+ e^- \rightarrow \pi^+ \pi^- \gamma$ and 
$e^+ e^- \rightarrow \mu^+ \mu^- \gamma$ are implemented in PHOKHARA. 
The results of \cite{rad6} include mass corrections needed in the limit
when the photon is emitted at small angles. The inclusion of such photons
in the radiative return cross-section is advantageous due to the 
enhanced rate for collinear emission.

The process $e^+ e^- \rightarrow f{\overline f} + n\gamma$ is also
implemented in the KK Monte Carlo.\cite{KKMC1,KKMC2}
In particular, the initial state radiative correction to the process 
$e^+ e^- \rightarrow \mu^+ \mu^- \gamma$ was calculated exactly\cite{virt1} at 
order ${\alpha^2}$. The KKMC was designed for high energy $e^+ e^-$ 
annihilation at LEP and LEP2, so the energies tested in ref.\ \cite{virt1}
were higher than for PHOKHARA,
but the initial state radiation was calculated to the same level of 
exactness in each case.
Previous results\cite{berends} (BVNB) and \cite{japan} (IN) 
for the virtual correction to 
initial state bremsstrahlung in this process have been compared to the 
results of \cite{virt1} (JMWY), but the results of BVNB are not
fully differential, and the results of IN do not include mass
corrections. 

The comparison of the virtual corrections of JMWY to those of \cite{rad5,rad6} 
(KR) is the closest presently available.  Since both are calculated with 
special attention to small photon angles, electron mass corrections are 
included in each expression, but by different means. This comparison
is a component of a Monte Carlo comparison, reported by 
S.\ Jadach in these proceedings,\cite{KKMC-PHOK} of 
the KKMC and PHOKHARA for muon pair or pion pair final states.
In that comparison, agreement to within
0.2\% was found for muon pair final states with pure initial state 
radiative corrections.  

\section{Comparison of Virtual Corrections}

In this note, we will compare an implementation of the initial state
virtual corrections of JMWY and KR directly in the 
context of the KKMC.  This comparison\cite{compare1,paris,beijing}
 is of particular interest because
the published matrix elements have very different forms, making an 
analytic comparison nontrivial.  This is most evident in the appearance of 
mass terms proportional to $(p_i\cdot k)^{-2}$  and $(p_i\cdot k)^{-3}$ in
the expressions of KR, and the absence of such terms in the 
expressions of JMWY, where $p_i$ is an incoming fermion momentum, and
$k$ is the emitted photon momentum.  We have verified
 that, in fact, all such
terms cancel exactly in the expressions of KR, leaving a leading
collinear factor of $(p_i\cdot k)^{-1}$. This cancellation should be
implemented analytically to obtain a stable evaluation in a MC program.  

We have also verified that in the massless limit,  the two expressions
for the virtual correction agree in the NLL limit, where the photon is 
taken to be collinear with an incoming fermion.\cite{beijing} This 
comparison also makes use of a careful expansion of the two expressions
in the collinear limit, which is needed to cancel apparent extra powers 
of $p_i\cdot k$ in the denominators in the expressions of KR.

The previously-available comparisons, refs. \cite{berends, japan},  
for the virtual photon correction to
ISR were earlier shown to agree with \cite{virt1} in the collinear limits. 
In the case of of \cite{berends}, this includes the mass corrections. 
However, these two comparisons are less complete.  In the case of \cite{berends}, the direction of the photon has been integrated, and in the case of 
\cite{japan}, the mass corrections needed for high precision 
in the collinear limit are not included. 

Since the four expressions JMWY, BVNB, IN and KR all agree analytically 
to NLL order in the massless limit, it is useful to compare the residual NNLL 
contribution after subtracting the common collinear limit of each expression.
In practice, this was done by calculating the YFS residual\cite{YFS1,YFS2}
$\overline{\beta}_1^{(2)}$ for single hard photon emission at order $\alpha^2$,
where a standard IR contribution has been subtracted. These
residuals are used in the implementation of the KKMC.\cite{KKMC1,KKMC2} 
In the collinear (NLL) limit, this residual can be related to 
$\overline{\beta}_1^{(1)}$ at order $\alpha$ (without the virtual 
photon)  via a form factor $f_{\rm NLL}$ such that 
\begin{equation}
\overline{\beta}_1^{(2)} = \overline{\beta}_1^{(1)} \left(1 + 
{\alpha\over 2\pi}\langle f_{\rm NLL}\rangle\right)
\end{equation}
with spin-averaged NLL form factor\cite{virt1}
\begin{eqnarray}
\label{NLL}
\langle f_{\rm NLL}\rangle &=& 2\left\{\ln\left({s\over m_e^2}\right) 
- 1\right\} + {r_1(1-r_1)\over 1 + (1-r_1)^2}
+ {r_2(1-r_2)\over 1 + (1-r_2)^2}
\nonumber\\
&+& 2\ln r_1 \ln (1-r_2)
+ 2\ln r_2 \ln (1-r_1)
- \ln^2 (1-r_1) - \ln^2 (1-r_2)
\nonumber\\
&+& 3\ln (1-r_1)
+ 3\ln (1-r_2)
+ 2\Sp(r_1)
+ 2\Sp(r_2) + \langle f_{\rm NLL}^{m}\rangle.
\end{eqnarray}
Here, $s = (p_1 + p_2)^2$, $r_i  = 2 p_i\cdot k/s$, and Sp$(z)$ is the 
Spence dilogarithm function. 
The NLL limit of the mass correction is taken to be \cite{virt1} 
\begin{eqnarray}
\label{eq:mass}
\langle f_{\rm NLL}^{m}\rangle &=& {2m_e^2\over s} 
        \left({r_1\over r_2} + {r_2\over r_1}\right)
{1 - r_1 - r_2 \over (1-r_1)^2 + (1-r_2)^2}
\nonumber\\
&\times & \left\{\langle f_{\rm NLL}^{m=0}\rangle 
+ \Big[\ln (1 - r_1) + \ln(1 - r_2) - 1\Big]
\ln\left({s\over m_e^2}\right) - {3\over 2} \ln(1 - r_1)
\right.\nonumber\\
& & \left. - {3\over 2} \ln (1 - r_2)
 + {1\over 2} \ln^2 (1 - r_1) + {1\over2} \ln^2(1 - r_2) + 1\right\}.
\end{eqnarray}
After subtracting this expression from each of the results,
we obtain the NNLL contribution to be compared. 

\begin{figure}[t]
\begin{center}
\hbox{\epsfig{file=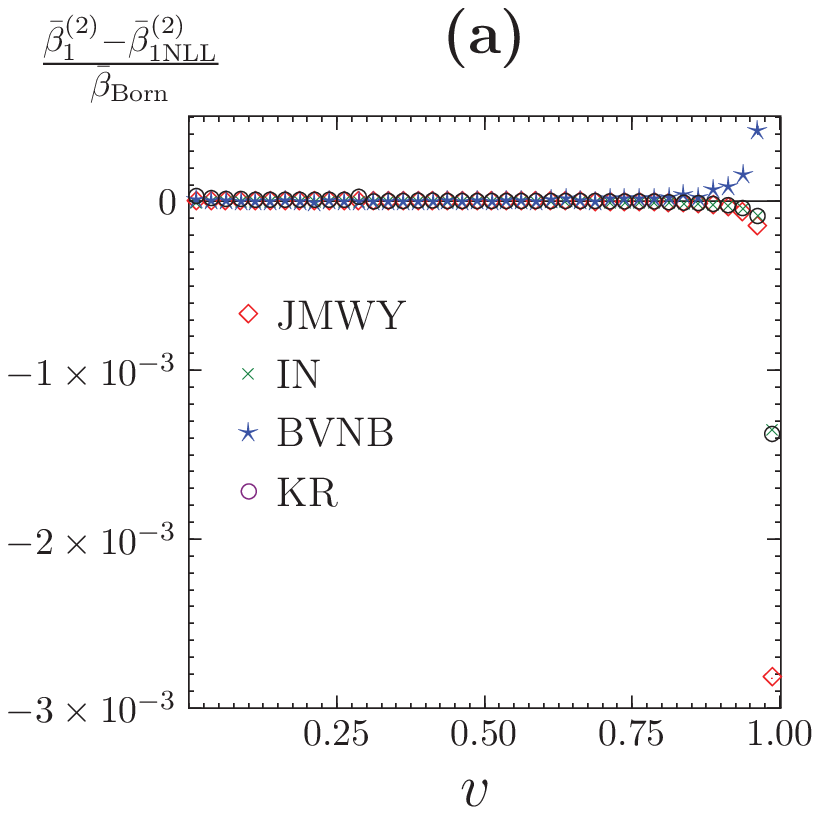,width=6.2cm}
\epsfig{file=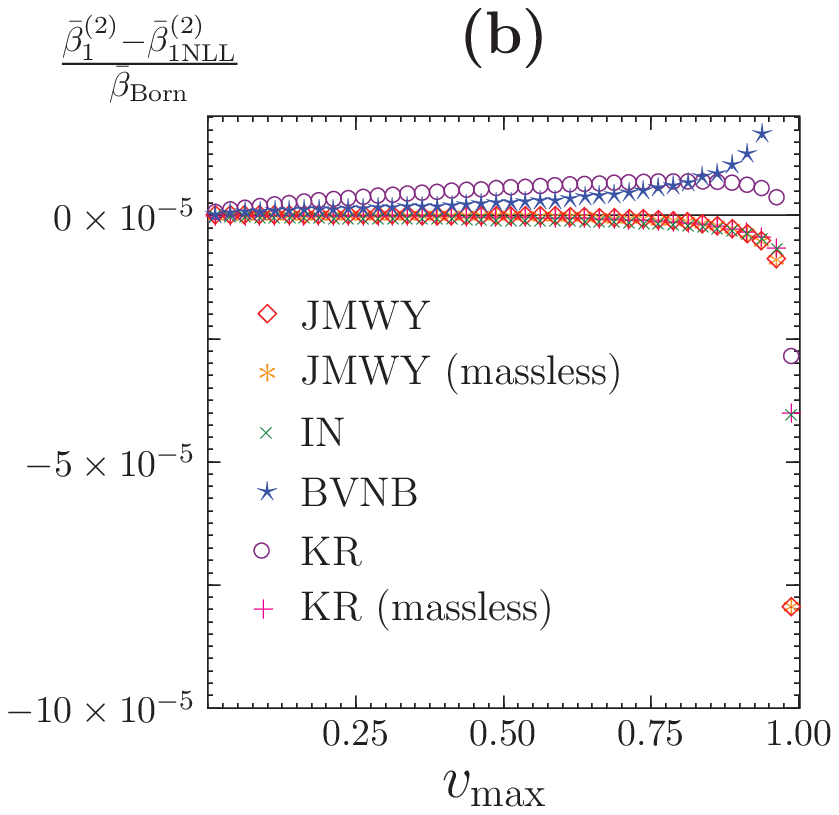,width=6.2cm}}
\end{center}
\caption{
Comparison of the NNLL contributions to ISR for muon pair production at
a CMS energy of 200 GeV.
The expressions are compared for $10^8$ events generated by the 
EEX3 option of the KK Monte Carlo as a function of the fraction $v$
of the beam energy radiated to the photon. The results is
in units of the Born cross-section.  Fig.\ (a) is differential in $v$,
and fig.\ (b) is integrated up to a cut $v_{\rm max}$. 
}
\label{fig:Fig1}
\end{figure}

The four expressions were implemented in the EEX3 option of the KK MC
(the YFS3ff generator),\cite{KKMC2}
and compared for muon pair production with pure initial state radiation (ISR).
Fig.\ 1 compares the NNLL  results for $10^8$ generated events. The 
CMS energy was chosen as 200 GeV to match the earlier comparisons in 
ref.\ \cite{virt1}. Fig.\ 1(a) shows a differential distribution in the 
photon energy fraction $v = 1 - Q^2/s$, where $Q$ is the effective CMS
momentum for the radiative return process.  In fig.\ 1(b), the cross-section has
been integrated up to an energy cut $v_{\rm max}$, as in the original
comparisons in ref.\ \cite{virt1}.  In order to compare both
the size of the NNLL effects and the mass corrections separately, fig.\ 1(b)
includes runs with and without the mass terms in the JMWY and KR
expressions.  Comparisons of the 
type in fig.\ 1(b) have been discussed in ref.\ \cite{beijing}, and with a 
different choice of NLL limit (equivalent up to collinear terms) in
refs.\ \cite{compare1,paris}.  

In the integrated cross-section, it is found that all of the results agree to within $0.5\times 10^{-5}$ in
units of the Born cross-section ($e^+ e^- \rightarrow \mu^+\mu^-$ without
radiation) up to a cut of $v_{\rm max} = 0.95$, except for the 
BVNB result, which is not fully differential in the photon momenta.  
For the last data point, with
a maximum $v = 0.975$, a larger departure is seen, with the difference 
between JMWY and KR results reaching $5.2\times 10^{-5}$ units of the 
Born cross-section.  

Differences in mass correction account for 
 $0.6\times 10^{-5}$ units of this difference.  This shows that
in spite of the apparent difference in the analytic expressions for the
mass terms, they are essentially equivalent. This is a nontrivial result,
since the mass corrections of KR were calculated by applying
FeynCalc\cite{FeynCalc} to the exact expression for the leptonic tensor,
while the mass corrections of JMWY were calculated using the methods
of ref.\ \cite{calcul}, after verifying that these methods reproduce the
exact mass corrections up to terms of order $m_e^2/s$ in the fully
integrated cross-section. This technique leads to a compact expression
for the essential contribution from the mass correction in the collinear
limit which can be evaluated without potential numerical difficulties which
can arise from higher powers of collinear factors in the denominators.

In the differential 
plot, fig.\ 1(a), the difference between the JMWY and KR results at $v = 0.975$
is found to be 1.4 per mil in units of the Born cross-section, with the
KR result in agreement with the IN result.  This difference is 
consistent with fig.\ 1(b), since it is due mostly to the last bin, which
includes 1/40 of the $v$ range. In fact, the difference between the 
results in the next next to the last bin, at $v = 0.95$, is only
$3\times 10^{-5}$ in units of the Born cross-section.  

\begin{figure}[b]
\begin{center}
\hbox{\epsfig{file=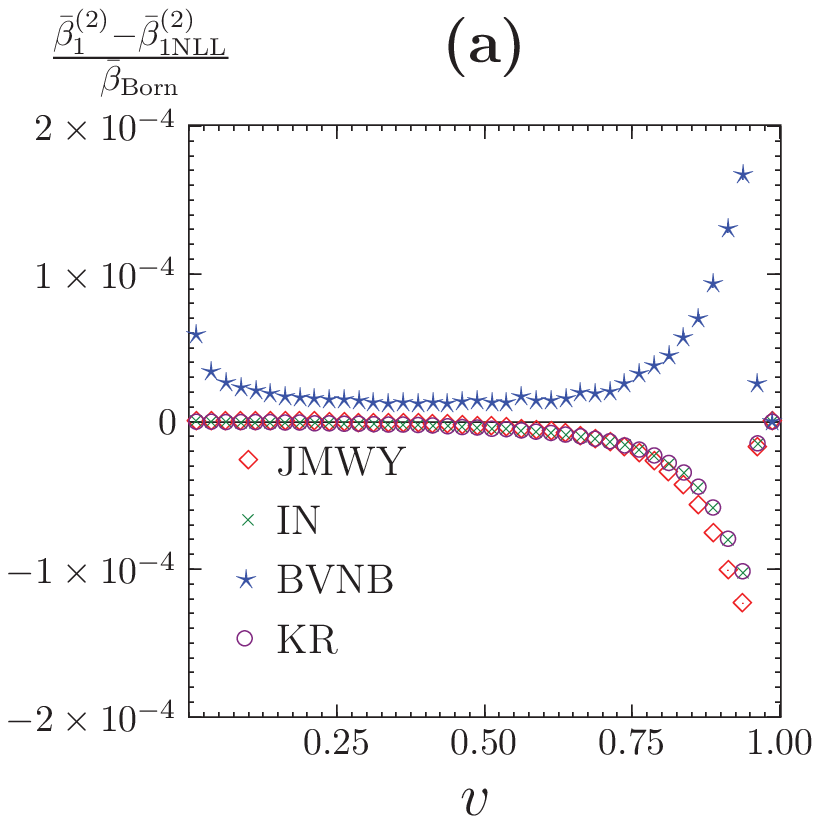,width=6.2cm}
\epsfig{file=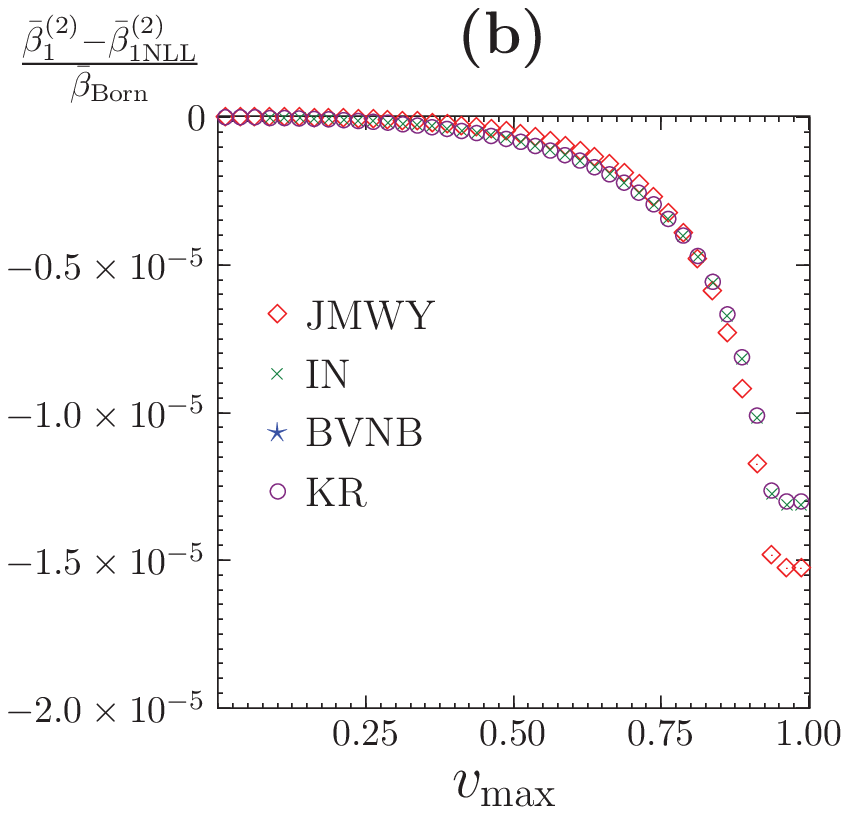,width=6.2cm}}
\end{center}
\caption{
Comparison of the NNLL contributions to ISR for muon pair production at
a CMS energy of 1.0 GeV.
The expressions are compared for $10^8$ events generated by the 
EEX3 option of the KK Monte Carlo as a function of the fraction $v$
of the beam energy radiated to the photon. The results is
in units of the Born cross-section.  Fig.\ (a) is differential in $v$,
and fig.\ (b) is integrated up to a cut $v_{\rm max}$. 
}
\label{fig:Fig2}
\end{figure}

The parameters for fig.\ 1 were chosen to match earlier comparisons\cite{virt1}
for the LEP2 final data analysis, when the KR result was not yet available. 
For radiative return, a low-energy comparison would be more appropriate.
In fig.\ 2, we have chosen 1 GeV as a representative $e^+e^-$ CMS energy. 
No cuts were applied on the fermion directions.  The BVNB result, which is
not differential in the photon directions, has a greater difference from
the other results at low CMS energy, but the JMWY and KR
results agree more closely, reaching a maximum
of $2\times 10^{-5}$ units
 of the Born cross-section in the differential distribution 2(a), 
and $2.3\times 10^{-6}$ in the integrated distribution 2(b). 
For $v < 0.85$, all of the differential results except BVNB agree to 
within $10^{-5}$ of the Born cross-section.

\section{Comparison of Virtual Corrections}

Our results show that the results of JMWY and KR for the virtual corrections
used in the calculation of radiative return agree to within $5\times 10^{-5}$
units of the Born cross-section for the full range of photon energies in 
the integrated distribution of fig.\ 1(b), or within 1.4 per mil in the 
differential distribution of fig.\ 1(a) at a CMS energy of 200 GeV. Over most
of the range of photon energies, the agreement is on the order of $10^{-5}$
or better.  

Excellent agreement is also found at a CMS energy of 1.0 GeV, an energy
scale more relevant for radiative return experiments at, for example, 
DA${\Phi}$NE.  Here, both the differential and integrated distributions in
fig.\ 2  show agreement on the order of $10^{-5}$ or better for the JMWY
and KR results over the entire range of photon energies. 

The comparison of the effect of mass corrections is of particular 
interest, since 
Differences in the treatment of mass corrections are the most
obvious distinction between the expressions of JMWY and KR at an analytic
level, but 
The MC results show that in fact, the difference between the mass corrections
is insignificant, less than $0.6\times 10^{-5}$ even in the large $v$ 
limit.

These results show that we have a clear understanding of the precision 
for the hard photon plus virtual photon contribution to the order 
$\alpha^2$ radiative correction to $f\overline{f}$ production, an 
important process not just in radiative return, but also in the 
final LEP2 data analysis and any anticipated future linear collider
physics.\cite{paris}

\vspace{12pt}
S.Y. would like to thank the organizers of the 2005 Cracow Epiphany 
Conference for their hospitality.  This work was supported in part 
by NATO grant PST.CLG.980342.

\label{references}

\end{document}